\documentclass[11pt]{article}
 \usepackage{amsmath, xypic, graphicx}
\usepackage{stmaryrd,xspace,amssymb}
\usepackage[active]{srcltx}
\usepackage{url}
\newcommand{\downset}{\ensuremath{\mathop{\downarrow\!}}}

\newcommand{\beq}{\begin{equation}}
\newcommand{\eeq}{\end{equation}}
\newcommand{\bea}{\begin{eqnarray}}
\newcommand{\eea}{\end{eqnarray}} \newcommand{\nn}{\nonumber}

\newcommand{\Sets}{\mbox{\textbf{Sets}}}

 \newcommand{\til}{\tilde}
\newcommand{\raw}{\rightarrow}

 \newcommand{\Raw}{\Rightarrow}

 \newcommand{\wed}{\wedge}

\newcommand{\x}{\times}

\newcommand{\inv}{^{-1}}

\newcommand{\er}{\eqref}

\newcommand{\lm}{\lambda} 
\newcommand{\rh}{\rho} \newcommand{\sg}{\sigma}
\newcommand{\Sg}{\Sigma}  
 \newcommand{\phv}{\varphi}

\newcommand{\CA}{{\mathcal A}}

   \newcommand{\CL}{{\mathcal L}}

\newcommand{\CO}{{\mathcal O}} \newcommand{\CP}{{\mathcal P}}

\newcommand{\C}{{\mathbb C}} 
 
 \newcommand{\R}{{\mathbb R}}
 
\newcommand{\alg}[1]{\ensuremath{#1}}
\newcommand{\functor}[1]{\ensuremath{\underline{#1}}}

\newcommand{\id}[1]{\ensuremath{\mathrm{id}}}

\newcommand{\Idl}{\ensuremath{\mathrm{Idl}}}

\newcommand{\Sh}{\ensuremath{\mathrm{Sh}}}

\newcommand{\context}{\ensuremath{\mathcal{C}}}

\newcommand{\sa}{\ensuremath{_{\mathrm{sa}}}}
\newcommand{\op}{\ensuremath{^{\mathrm{op}}}}

\newcommand{\prop}[1]{\ensuremath{\mbox{\texttt{#1}}}}

\newcommand{\uA}{\underline{A}}

\renewcommand{\CA}{\mathcal{C}(A)}

\newcommand{\ulS}{\functor{\Sg}}

\renewcommand{\CA}{\context(\alg{A})}
\newtheorem{theorem}{Theorem}

\newtheorem{proposition}[theorem]{Proposition}

{ \begin{trivlist}%
  \item[\hskip \labelsep {\bfseries #1}]%
}%
{ \end{trivlist}%
}

\topmargin = - 0.5 cm
\newcommand{\turndown}[1]{%
  \rotatebox[origin=c]{270}{\ensuremath#1}}
\newcommand{\twoheaddownarrow}{\turndown{\twoheadrightarrow}}
\begin{document}
\title{The Gelfand spectrum of a noncommutative C*-algebra: \\  a topos-theoretic approach}
\author{Chris Heunen\footnote{Computing Laboratory, Oxford University, Wolfson Building, Parks Road, Oxford OX1 3QD, U.K. 
Email: \texttt{heunen@comlab.ox.ac.uk}.
Supported by N.W.O.\ through a Rubicon grant.}\and 
  Nicolaas P. Landsman\footnote{Radboud Universiteit Nijmegen,
    Institute for Mathematics, Astrophysics, and Particle Physics,
   Heyendaalseweg 135, 6525 {\sc aj nijmegen, the netherlands}.
   Email: \texttt{landsman@math.ru.nl}.
   } \and
  Bas Spitters\footnote{Radboud Universiteit Nijmegen,   Institute for 
 Computer and Information Science,  Heyendaalseweg 135, 6525
 {\sc aj nijmegen, the netherlands}.   Email: \texttt{spitters@cs.ru.nl}. Supported by N.W.O.\ through the {\sc diamant} cluster.}\and
    Sander Wolters\footnote{Radboud Universiteit Nijmegen,
    Institute for Mathematics, Astrophysics, and Particle Physics,
   Heyendaalseweg 135, 6525  {\sc aj nijmegen, the netherlands}.    Email: \texttt{s.wolters@math.ru.nl}.
    Supported by N.W.O.\ through project 613.000.811.} }
\maketitle
\vspace*{-0.75cm}
\begin{center}{\it
Dedicated to Alan Carey, on the occasion of his 60$\,^{th}$ birthday}
\end{center}
\smallskip
\begin{abstract}
We compare two influential ways of defining a generalized notion of space. The first, inspired by Gelfand duality,
states that the category of `noncommutative spaces' is the opposite of the category of C*-algebras.
The second, loosely generalising Stone duality, maintains that  the category of `pointfree spaces' is the opposite of the category of frames
(i.e., complete lattices in which the meet distributes over arbitrary joins). One possible relationship between these two notions of space 
was unearthed by Banaschewski and Mulvey  [``A globalisation of the {G}elfand duality theorem'', Annals of Pure and Applied Logic 137, 62--103 (2006)], who proved a constructive version of Gelfand duality in which the Gelfand spectrum of a commutative C*-algebra comes out as a pointfree space.
 Being constructive, this result applies in arbitrary toposes (with natural numbers objects, so that internal C*-algebras can be defined). 

Earlier work by the first three authors [``A topos for algebraic quantum theory'', Communications in Mathematical Physics 291, 63--110 (2009)],  shows how a noncommutative C*-algebra gives rise to a commutative one internal to a certain sheaf topos. The latter, then, has a constructive Gelfand spectrum, also internal to the topos in question. After a brief review of this work, we compute the so-called external description of this internal spectrum, which in principle is a fibered pointfree space in the familiar topos \Sets\  of sets and functions. 
However, we obtain the external spectrum as a fibered topological space in the usual sense.
This leads to an explicit Gelfand transform, as well as to a topological reinterpretation  of the Kochen--Specker Theorem of quantum mechanics [``The problem of hidden variables in quantum mechanics'',  Journal of Mathematics and Mechanics 17, 59--87 (1967)], which supplements the remarkable topos-theoretic version of this theorem due to  Butterfield and Isham [``A topos perspective on the {K}ochen-{S}pecker theorem", 
International Journal of Theoretical Physics  37, 2669-2733 (1998)].
\end{abstract}
\newpage
\section{Generalized spaces}
Gelfand Duality is the categorical equivalence
\begin{equation}
\mbox{compact Hausdorff spaces} \simeq (\mbox{unital commutative C*-algebras})^{op}, \label{GD}
\end{equation}
where the choice of arrows in both categories is  implicit (but obvious, i.e., continuous maps and unital $\mbox{}^*$-homomorphisms, respectively). For simplicity, we restrict ourselves to the compact/unital case. 
Furthermore, given a  category $\mathsf{C}$,  the opposite category 
$\mathsf{C}^{op}$ has the same objects as $\mathsf{C}$, but has all arrows reversed. The relevant functors implementing the equivalence \er{GD} are, of course, $C:X\mapsto C(X)\equiv C(X,\C)$ from left to right, with pullback on arrows,
and $\Sg: A\mapsto \Sg(A)$ from right to left, where $\Sg(A)$ is the Gelfand spectrum of $A$ (realized, e.g., as the space of unital multiplicative linear maps $A\raw\C$ equipped with the relative weak$\mbox{}^*$-topology), and similarly  pullback on arrows.

Subsequently, there are (at least) two possible directions to take.
\begin{enumerate}
\item 
 The modern approach is to literally take the 
quantum jump of  \emph{defining} the category of    `noncommutative spaces' up to equivalence by
\begin{equation}
\mbox{noncommutative spaces} \simeq (\mbox{C*-algebras})^{op}. \label{GD2}
\end{equation}
Here a major surprise arises, which is quite unexpected from the categorical setting: according to the (second) Gelfand--Naimark Theorem, a noncommutative space acts as an operator algebra on some Hilbert space. It is the combination of this Hilbert space setting deriving from the right-hand side of \er{GD2} and the call for
geometrical and topological techniques  - adapted to the noncommutative setting -  coming from the left-hand side that gives noncommutative geometry its strength \cite{Connes,ConnesMarcolli}.
\item
More traditionally, one may attempt to generalize the notion of Gelfand duality to noncommutative C*-algebras $A$. There have been many such attempts, which may be grouped according to the specific notion of a Gelfand spectrum that is used. For example, in the 
Dauns--Hofmann Theorem \cite{DaunsHofmann,FellDoran2, Mulvey80} the Gelfand spectrum of $A$ is taken to be the Gelfand spectrum of its centre $Z(A)$, on which $A$ is realized as a sheaf.  Akemann, on the other hand, used the space of maximal left ideals of $A$, but needed to generalize the notions of topology and continuity \cite{Akemann}. Shultz used the pure state space of $A$, equipped with the structure of a transition probability \cite{Shultz}, later refined so as to make the noncommutative Gelfand spectrum a so-called Poisson space with a transition probability  \cite{PSTP,landsman98}.
See also \cite{CIR,Woro}, etcetera.  In all  cases, the point is to realize $A$ in a way that resembles a space of complex-valued continuous functions as much as possible. 
\end{enumerate}
Ultimately, what lies behind both directions is the success of Gelfand duality in capturing (compact Hausdorff) spaces algebraically. What is slightly unnatural, though, is that this capturing should involve the complex (or, for that matter, the real) numbers in a fundamental way. This may be avoided in an order-theoretic approach, as follows \cite{johnstone82}, \cite[Ch.\ IX]{maclanemoerdijk92}. Instead of the passage $X\mapsto C(X)$ from spaces to complex algebras, we take $X\mapsto \CO(X)$, where $\CO(X)$ is just the topology of $X$ in the defining sense of its collection of open sets.
This has a natural lattice structure under inclusion, and in fact defines a highly structured kind of lattice known as a {\it frame}. This is  a complete distributive lattice such that
$x\wedge \bigvee_{\lambda}y_{\lambda}=\bigvee_{\lambda}x\wedge
y_{\lambda}$ for arbitrary families $\{y_{\lambda}\}$ (and not just for finite ones, in which case the said property follows from the definition of a distributive lattice).  Indeed,
 $\CO(X)$ is a frame with $U\leqslant V$ if $U\subseteq V$.
A {\it frame homomorphism} preserves finite meets and arbitrary
joins; this leads to the category of frames and frame homomorphisms.

In order to have an equivalence like \er{GD}, we need to cut down both the category of spaces and the category of frames. To do so, we first define a \emph{point} of a frame $F$ as a frame map $p^*:F\raw\{0,1\}$, where as a frame $\{0,1\}$ is identified with $\CO(*)$, i.e., the topology of a space with a point (so that we identify $0$ with $\emptyset$ and $1$ with $*$).  In fact, if $F=\CO(X)$, then any point $p\in X$ defines a point of $F$ by $p^*=p\inv$ (that is, $p^*(U)=1$ iff $p\in U$).
Using this concept, the set $\mathrm{Pt}(F)$ of points of a frame $F$ may be topologized in a natural way, by declaring its opens to be
the sets of the form
$\mathrm{Pt}(U)=\{p^*\in \mathrm{Pt}(F)\mid p^*(U)=1\}$,
 where  $U\in F$.  We say that a frame $F$ is {\it spatial}
if it is isomorphic (in the category of frames) to
$\CO(\mathrm{Pt}(F))$.
On the other hand, a topological space
$X$ is called {\it sober} if it is homeomorphic to $\mathrm{Pt}(\CO(X))$. Given these definitions, it is almost tautological that
\begin{equation}
\mbox{sober spaces} \simeq (\mbox{spatial frames})^{op}, \label{SD}
\end{equation}
where the equivalence is given by  $\CO$ and Pt (seen as functors).\footnote{\label{sfn} Though \er{SD} is true almost by definition, the nontrivial statement of Stone duality, i.e.,
Stone spaces $\simeq (\mbox{Boolean lattices})^{op}$,  is actually a special case of \er{SD}. The nontrivial observation - apart from the fact that Hausdorffness implies soberness - is that although Stone spaces form a subcategory of sober spaces, Boolean lattices are not a subcategory of frames (for one thing, a Boolean lattice need not be complete). Hence a special manoeuvre is needed to embed Boolean lattices in frames, which is done through the so-called \emph{ideal completion} $L\mapsto \Idl(L)$; this is the
collection of nonempty lower closed subsets $I\subset  L$ such that $x,y\in I$ implies $x\vee y\in
I$, ordered by inclusion \cite[p.59]{johnstone82}.  A Stone space $X$ then defines the Boolean lattice $\CO_c(X)$ of clopen subsets of $X$, whose ideal completion is the topology $\CO(X)$; conversely,
a Boolean lattice $L$ defines a Stone space $X=\mathrm{Pt}(\Idl(L))$, with $\CO(X)\cong \Idl(L)$.}
Let us note the following, however. It is easily shown that a frame $F$ is spatial iff $F\cong \CO(X)$ for \emph{some} space $X$, not necessarily sober - in fact, we will later encounter an example of exactly this situation. In that case, following \cite{johnstone82}, we may call 
\begin{equation}
X^S=\mathrm{Pt}(\CO(X)), \label{soberification}
\end{equation}
which is necessarily sober,  the \emph{soberification} of $X$ (if $X$ is already sober, one has $X^S\cong X$).
 This construction may be compared to the passage from a compact non-Hausdorff space $X$ to  its \emph{Hausdorffication}  
 \begin{equation}
X^H=\Sg(C(X)). \label{Ha}
\end{equation}

Now recall that the step from \er{GD} to \er{GD2} introduced a certain generalization of the concept of space by omitting the qualifier ``unital commutative'' in the characterization of spaces in the right-hand side of  \er{GD}. Analogously, we may omit the qualifier ``spatial''
in the right-hand side of  \er{SD}, hoping to arrive at a \emph{different} generalized notion of space. 
Following  \cite{johnstone82,joyaltierney84, maclanemoerdijk92}, we therefore
 write
\begin{equation}
\mbox{pointfree spaces} \simeq (\mbox{frames})^{op},\label{SD3}
\end{equation}
which, like \er{GD2}, is no longer a duality \emph{theorem}, but a statement of  the \emph{definition} of the category of `pointfree spaces' (also known as \emph{locales}). This  definition comes with a curious piece of notation: 
\emph{any} frame is written $\CO(X)$, whether or not it is spatial, and the corresponding pointfree space is written as $X$. Furthermore, the symbol $C(X,Y)$ denotes the object (in whatever category the frames are defined) of frame maps from $\CO(Y)$ to $\CO(X)$; a `continuous' map $f:X\raw Y$ is nothing but a frame map from $\CO(Y)$ to $\CO(X)$, which tends to be written as $f^*$ or $f\inv$. 
 This notation is partly motivated by the case where  $\CO(X)$ are $\CO(Y)$ actually the topologies of sober spaces $X$ and $Y$, respectively, for in that case it can be shown (nonconstructively) that
 any frame map $f^*:\CO(Y)\raw\CO(X)$ is of the form $f^*=f\inv$ for a continuous map $f:X\raw Y$ in the usual sense. 
 
The surprising role of Hilbert spaces in the theory of noncommutative spaces has a counterpart for pointfree spaces: these turn out to be related to \emph{logic}, especially to \emph{intuitionistic propositional logic}.\footnote{Perhaps this is less surprising in view of Stone duality and  the well-known connection between Boolean lattices and classical propositional logic.} 
Indeed, a frame is a  {\it complete Heyting algebra}, where a  Heyting algebra
is a distributive lattice $\CL$ with a map $\raw:\CL\x\CL\raw\CL$
satisfying $x\leqslant (y\raw z)$
iff $x\wed y\leqslant z$, called implication  \cite{goldblatt84,maclanemoerdijk92,Vic:LocTopSp}. 
Unlike in a Boolean lattice, negation is now a derived notion, defined by $\neg x=(x\raw\perp)$. 
 Every Boolean algebra is a Heyting algebra, but not {\it vice versa}; in fact, a Heyting
algebra is Boolean iff $\neg\neg x=x$ for all $x$, which is the case
iff $\neg x\vee x=\top$  for all $x$; not necessarily granting this is the essence of intuitionistic logic. The point, then, is that
a complete Heyting algebra is essentially the same thing as a
frame, for  in a frame one may define $y\raw z=\bigvee\{x\mid x\wed y\leqslant z\}$, and
conversely, the infinite distributivity law in a frame is
automatically satisfied in a Heyting algebra.

In principle, noncommutative spaces and pointfree spaces (i.e., locales) appear to be totally different generalizations of the notion of a topological space. However, a close connection arises if we return to Gelfand duality. 
To explain this, note that the usual proofs of Gelfand duality are nonconstructive; for example, if the Gelfand spectrum is realized as the maximal ideal space of $A$, one needs Zorn's Lemma. However, a typical situation in constructive mathematics now arises: Gelfand duality is nonconstructively equivalent to a result that is constructively valid (that is, provable without using the axiom of choice or the exclusion of the middle third) \cite{banaschewskimulvey00b, banaschewskimulvey00a,banaschewskimulvey06, coquand05,CS}.
Hence the constructive version of the key ingredient of classical Gelfand duality, namely the isomorphism
\begin{equation}
A\cong C(\Sg(A),\C) \label{CGD}
\end{equation}
of commutative C*-algebras, is formally \emph{the
very same statement}, but now \emph{reinterpreted} according to the notation for frame maps just explained. Thus
the Gelfand spectrum $\Sg(A)$ and the complex numbers $\C$ are now objects of the category of
 pointfree spaces, i.e., they are really frames $\CO(\Sg(A))$ and $\CO(\C)$, which are not necessarily spatial,\footnote{
 Technically, $\CO(\Sg(A))$ is
required to be compact and completely regular \cite{banaschewskimulvey06}, which are frame-theoretic properties replacing the combination compact Hausdorff for topological spaces \cite{johnstone82}.} and $C(\Sg(A),\C)$ denotes the object (in the ambient category) of frame maps from $\CO(\C)$ to $\CO(\Sg(A))$.

The choice between the constructive version of Gelfand duality (in terms of pointfree spaces) and its familiar nonconstructive counterpart  (in terms of topological spaces) 
is not a matter of philosophical taste. In set theory, the usual version is perfectly acceptable to us.
 The point is that constructive Gelfand duality holds in arbitrary toposes  (with natural numbers objects, so that internal C*-algebras can be defined).\footnote{We refer to \cite{johnstone02a, johnstone02b} for an encyclopaedic treatment of topos theory, to
 \cite{maclanemoerdijk92}, \cite{borceux3}, or \cite{goldblatt84} for (complementary) book-length introductions, to \cite{Vic:LocTopSp} for a shorter treatment, and finally  to the appendix of \cite{CHLS} for a very brief survey of exactly what is needed below. The notion of a C*-algebra in a topos with natural numbers object, including the statement and proof of Gelfand duality in the commutative case, is due to Banaschewski and Mulvey \cite{banaschewskimulvey06}. See also \cite{HLS} for a review of this theory, including a reformulation along the lines of \cite{coquand05,CS}.}
    \section{Internal Gelfand spectrum}
In order to define Gelfand spectra for noncommutative C*-algebras, 
 we proceed as follows \cite{HLS}.
Let $A$ be a unital C*-algebra, and let $\CA$ be the poset of unital \emph{commutative} C*-subalgebras of $A$ (ordered by set-theoretic inclusion), equipped with the Alexandrov topology.\footnote{The open sets $U$ of the Alexandrov topology on a poset $P$.
 are  the upward closed sets  (if $x\in U$ and $x\leq y$, then $y\in U$). 
 The sets  $U_x=\mathop{\uparrow}\!  x=\{y\in P\mid y\geq x\}$, $x\in P$,  form a basis of the Alexandrov topology. } Thus we have the topos $\Sh(\CA)$ of sheaves on $\CA$.
 We now define a specific sheaf $\uA$ on $\CA$ by\footnote{This formula defines  $\uA$ on the basic opens $U_C=\mathop{\uparrow}\! C$ of $\CA$ in the Alexandrov topology. On an arbitrary open $U=\bigcup_{C\in\Gamma} U_C$, the sheaf property gives $\uA(U)= \lim_{C\in\Gamma} \uA(U_C)$. Under the identification of $\Sh(P)$ with $\Sets^{P}$ (where the poset $P$ is seen as a category in the usual way) through the correspondence $F(\mathop{\uparrow}\! x)\leftrightarrow F(x)$ \cite{goldblatt84}, the sheaf $\uA$ corresponds to the tautological functor $C\mapsto C$ in  $\Sets^{\CA}$. }
\begin{equation}
\uA(\mathop{\uparrow}\! C)=C; \:\: C\in \CA; \label{Bohr}
\end{equation}
if $C\subseteq D$, then $\mathop{\uparrow}\! D\subseteq \mathop{\uparrow}\! C$, and the map $\uA(\mathop{\uparrow}\! C)\raw \uA(\mathop{\uparrow}\! D)$, i.e., $C\raw D$, is simply given by inclusion. 
This sheaf turns out to be a \emph{commutative} C*-algebra $\uA$ in $\Sh(\CA)$ under natural operations, so that it has an internal Gelfand spectrum $\ulS(\uA)$. With $A$ fixed, we will henceforth simply call this spectrum $\ulS$; it is a pointfree space in the topos $\Sh(\CA)$.\footnote{The functorial properties of the map $A\mapsto \ulS(\uA)$, as well as of the map $A\mapsto\Sg(A)$ to be introduced below, have  been studied in \cite{BBH}.}

The explicit computation of $\ulS$ was initiated in \cite{HLS}, and was completed in \cite{SW}. To state the result (i.e., Theorem \ref{t1} below), topologize the disjoint union
\begin{equation}
\Sg=\coprod_{C\in\CA} \Sg(C),
\end{equation}
where $\Sg(C)$ is the usual Gelfand spectrum of $C\in\CA$ (i.e., the set of pure states or characters on $C$ with the relative weak $\mbox{}^*$-topology)
by saying that  $\mathcal{U}\in\CO(\Sg)$ iff the following two conditions are satisfied for all $C\in\CA$ (with the notation  $\mathcal{U}_C\equiv \mathcal{U}\cap\Sg(C)$):
\begin{enumerate}
\item  $\mathcal{U}_C\in\CO(\Sg(C))$.
\item For all $D\supseteq C$, if $\lm\in \mathcal{U}_C$ and $\lm'\in \Sg(D)$ such that $\lm'_{|C}=\lm$, then $\lm'\in \mathcal{U}_D$.
\end{enumerate}
For each $U\in\CO(\CA)$, we also introduce the space
 \begin{equation}
 \Sg_U=\coprod_{C\in U} \Sg(C), 
\end{equation}
with relative topology  inherited from $\Sg$. We then have:
\begin{theorem}\label{t1}
Let $A$ be a unital C*-algebra $A$. The frame $\CO(\ulS)$ in $\Sh(\CA)$ that underlies the internal Gelfand spectrum
$\ulS\equiv \ulS(\uA)$ of the internal commutative C*-algebra $\uA$ defined by \er{Bohr}
 is given by the sheaf
\begin{equation}
\CO(\ulS):U\mapsto \CO(\Sg_U), \label{sander}
\end{equation}
where $U\in\CO(\CA)$; if $U\subseteq V$, the map $\CO(\Sg_V)\raw \CO(\Sg_U)$  is given by $\mathcal{U}\mapsto \mathcal{U}\cap \Sg_U$.
\end{theorem}
The proof of this theorem is quite lengthy, requiring familiarity with constructive mathematics, as well as with the closely related technique of  internal reasoning in topos theory. Besides the general theory of internal Gelfand duality in Grothendieck toposes due to Banaschewski and  Mulvey  \cite{banaschewskimulvey06} looming in the background,
the proof of Theorem \ref{t1} consist of  three main steps:
\begin{enumerate}
\item The lattice-theoretic description of general constructive  Gelfand spectra \cite{coquand05,CS};
\item The specific application of this description to the commutative C*-algebra $\uA$ in the topos $\Sh(\CA)$ \cite{HLS};
\item The insight that this application yields the explicit form \er{sander} \cite{S2010,SW}.
\end{enumerate}
We now give a summary of these steps, referring to the papers just cited for further details.\footnote{In fact, the third step can be carried out in two rather different ways, of which the approach of \cite{SW} is easier to explain to operator algebraists. Hence in what follows we use the latter. The techniques in \cite{S2010} will be further explored in future work in collaboration with Steven Vickers, whom we wish to thank for his insightful comments on an earlier version of this paper. } In what follows, $A$ is a commutative C*-algebra with unit
in some topos (with natural numbers object), whereas $C$ is  a commutative C*-algebra with unit in the usual sense, i.e., in the topos  \Sets\ of sets and functions.  

\medskip

\noindent 1. As already mentioned, the constructive approach to Gelfand duality emphasizes the frame $\CO(\Sg)$ rather than the set $\Sg\equiv\Sg(A)$. To construct $\CO(\Sg)$, take the usual  positive cone 
 $A^+:=\{a \in A\sa \mid a \geq 0\}$ of $A$  (where $A_{\mathrm{sa}}$ is the selfadjoint part of $A$), and 
define $a \preccurlyeq
b$ iff there exists $n\in\mathbb{N}$ such that $a\leqslant
n b$. Define $a \approx b$ iff $a\preccurlyeq b$ and $b\preccurlyeq a$. The
lattice operations on $A\sa$ (defined with respect to the usual partial order $\leq$) respect $\approx$ and hence $L_A=A^+/\approx$ is a lattice under the descent of $\leq$ to the quotient, which we denote by $\leqslant$.

If $A$ is finite dimensional, the constructive Gelfand spectrum of $A$ is simply (isomorphic to)  the ideal completion $\Idl(L_A)$ of $L_A$ (cf.\ footnote \ref{sfn}). In general, one needs a refinement of this construction. First, define a surjective map $A\sa\raw L_A$,
$a\mapsto \prop{D}_a\equiv [a^+]$, where $a=a^+-a^-$, $a^{\pm}\in A^+$, and $[a^+]$ is the equivalence class of $a^+$ in $L_A$ with respect to $\approx$. Second, write  $\prop{D}_b\ll
\prop{D}_a$ iff $\prop{D}_b\leqslant \prop{D}_{a-q}$  for some $q >0$, $q\in\mathbb{Q}$. Third, we refine the  down-set
 $\downset \prop{D}_a=\{\prop{D}_b\in L_A\mid \prop{D}_b\leqslant \prop{D}_a\}$ to $\twoheaddownarrow \prop{D}_a=\{\prop{D}_b\in L_A\mid \prop{D}_b\ll \prop{D}_a\}$, and
declare an ideal $I\in \mathrm{Idl}(L_A)$ to be {\it  regular} if $I\supseteq \twoheaddownarrow \prop{D}_a$ for some $\prop{D}_a\in L_A$ implies $\prop{D}_a\in I$ (in other words, if $\prop{D}_b\in I$ for all $\prop{D}_b\ll \prop{D}_a$, then $\prop{D}_a\in I$).
This yields the frame $\mathrm{RIdl}(L_A)$  of \emph{regular} ideals of $L_A$, ordered by inclusion (like $\Idl(L_A)$, of which $\mathrm{RIdl}(L_A)$ is a subframe). 
The  constructive Gelfand spectrum of $A$, then,  turns out to be (isomorphic to) just this subframe, that is,
\begin{equation}
\CO(\Sg(A))\cong \mathrm{RIdl}(L_A). \label{eq12}
\end{equation}

There is a natural map $\til{f}_A:L_A\raw \mathrm{Idl}(L_A)$,  $\prop{D}_a\mapsto \downset \prop{D}_a$, which may be refined to a map 
$f_A:L_A\raw \mathrm{RIdl}(L_A)$
 that sends $\prop{D}_a$ to the smallest regular ideal containing $\til{f}_A(\prop{D}_a)=\downset \prop{D}_a$;
explicitly, one has $
f_A(\prop{D}_a)=\{\prop{D}_c\in L_A\mid \prop{D}_b\ll \prop{D}_c \Raw \prop{D}_b\leqslant\prop{D}_a,  \prop{D}_b\in L_A\}$.

If one thinks of $\CO(\Sg)$ as the `topology' of the Gelfand spectrum (in the appropriate pointfree sense), the `opens' $f_A(\prop{D}_a)$
(or, less accurately, the elements $\prop{D}_a$ of $L_A$ themselves), comprise `basic opens' for the topology, in terms of which general `opens' $U\in \mathrm{RIdl}(L_A)$ may be expressed as $U=\bigvee\{f_A(\prop{D}_a)\mid \prop{D}_a\in L_A, f_A(\prop{D}_a)\leq U\}$.

Applying this to ordinary unital commutative C*-algebras $C$, one finds that the frame $\CO(\Sg)$ is spatial, being related to the usual Gelfand topology $\CO(\Sg(C))$
by the frame isomorphism $\mathrm{RIdl}(L_C) \raw \CO(\Sg(C))$ that on basic opens is given by
$$f_C(\prop{D}_a) \mapsto \mathcal{D}_a\equiv \{\phv\in\Sg(C)\mid \phv(a)>0\},\:\: a\in C\sa.$$
In particular, 
the map
\begin{equation}
\prop{f}_C: L_C  \raw \CO(\Sg(C)) , \:\:
 \prop{D}_a\mapsto \mathcal{D}_a \label{eq14}
\end{equation}
is well defined (i.e., independent of the choice of $a$); cf.\ \cite[Lemma 2.14]{SW}.

\medskip

\noindent 
2.  Internalizing the above construction of $\CO(\Sg)$ to the topos $\Sh(\CA)$ and applying it to the internal C*-algebra $\uA$
first yields a lattice $\underline{L}_{\uA}$ in $\Sh(\CA)$, given by \cite[Theorem 20]{HLS}
\begin{equation}
\underline{L}_{\uA}(\mathop{\uparrow}\! C)=L_C. \label{LC}
\end{equation}
Interpreting $\mathrm{RIdl}$
 in the topos $\Sh(\CA)$ through Kripke--Joyal semantics  \cite{maclanemoerdijk92} then shows that 
the internal frame $\underline{ \mathrm{RIdl}}(\underline{L}_{\uA})$
 in $\Sh(\CA)$ is given by  the sheaf (cf.\ \cite[Theorem 29]{HLS})
\begin{equation}
\underline{ \mathrm{RIdl}}(\underline{L}_{\uA}): U\mapsto \{\underline{F}\in \mathrm{Sub}(\underline{L}_{\uA|U})\mid \underline{F}(\mathop{\uparrow}\! C)\in \mathrm{RIdl}(L_C)\mbox{ for all } C\in U\}. \label{FC}
\end{equation}
Here $\underline{L}_{\uA|U}:\CO(U)\op\raw\Sets$ denotes the restriction of the sheaf  $\underline{L}_{\uA}: \CO(\CA)\op\raw\Sets$ to $\CO(U)$, where
$U\in \CO(\CA)$, 
and $\mathrm{Sub}(\underline{L}_{\uA|U})$ is the set of subsheaves of $\underline{L}_{\uA|U}$; note that $\underline{F}(\mathop{\uparrow}\! C)\subseteq L_C$ by \er{LC}, so that  $ \underline{F}(\mathop{\uparrow}\! C)\in \mathrm{RIdl}(L_C)$ in \er{FC} is well defined. If $U\subseteq V$, then the map $\underline{ \mathrm{RIdl}}(\underline{L}_{\uA})(V)\raw\underline{ \mathrm{RIdl}}(\underline{L}_{\uA})(U)$ is given by restricting
$\underline{F}\in \mathrm{Sub}(\underline{L}_{\uA|V})$ to $\CO(U)$. 

\medskip

\noindent 
3. To prove \er{FC},  the transformation
$\theta:\underline{ \mathrm{RIdl}}(\underline{L}_{\uA})\raw\CO(\ulS)$ defined by its components
\begin{eqnarray}
\theta_U&:& \{\underline{F}\in \mathrm{Sub}(\underline{L}_{\uA|U})\mid \underline{F}(\mathop{\uparrow}\! C)\in \mathrm{RIdl}(L_C)\mbox{ for all } C\in U\}\raw\CO(\Sg_U); \nn\\
& & \underline{F}\mapsto \coprod_{C\in U}\bigcup_{\prop{D}_a\in \underline{F}(\mathop{\uparrow}\! C)}\mathcal{D}_a,
\end{eqnarray}
can be shown to be a natural isomorphism (since $\underline{ \mathrm{RIdl}}(\underline{L}_{\uA})$ and $\CO(\ulS)$ are internal frames, it suffices to prove that $\theta_{\CA}$ is an isomorphism of frames in \Sets; cf.\ \cite[Theorem 2.17]{SW}). Note that $\theta_U(\underline{F})$ indeed lies in $\CO(\Sg_U)$ by 
the property $\rho_{DC}^{-1}\circ \prop{f}_C= \prop{f}_D\circ \iota_{CD}$ for all $C\subseteq D$, $C,D\in\CA$, where  $\rho_{DC}^{-1}:\CO(\Sg(C))\raw \CO(\Sg(D))$ is the  inverse image map of the restriction
 $\rho_{DC}:\Sg(D)\raw\Sg(C)$, $\lm\mapsto\lm_{|C}$,
and  $\iota_{CD}:L_C\raw L_D$ is the obvious embedding $\prop{D}_a\mapsto \prop{D}_a$ (where $a\in C$ in the first  $\prop{D}_a$ and $a\in D$ in the second).
  \hfill Q.E.D.

\medskip

We illustrate Theorem \ref{t1} for $A=M_n(\C)$, i.e., the $n\x n$ complex matrices. We then have a frame isomorphism 
$\CO(\Sg(C))\cong \CP(C)$ for any $C\in\CA$ \cite{CHLS}, where $\CP(C)$ is the projection lattice of $C$ (and similarly,  $\CP(A)$ below is the projection lattice of $A$).
Hence
\begin{equation}
\CO(\Sg)\cong \{S:\mathcal{C}(A) \raw \CP(A)\mid {S}(C)\in \CP(C),\, 
{S}(C)\leq {S}(D)\:\mbox{if}\: C\subseteq D\}, \label{eerste}
\end{equation}
where 
 the right-hand side is equipped with the pointwise partial order $\leq$ with respect to
the usual partial ordering $\leqslant$ of projections, i.e., $S\leq T$ iff  $S(C)\leqslant T(C)$ for all  $C\in\CA$. To obtain \er{eerste} we identify 
$\mathcal{U}=\coprod_{C\in\CA} \mathcal{U}_C$ as an element of $\CO(\Sg)$
 with $S:\mathcal{C}(A) \raw \CP(A)$ on the right-hand side of \er{eerste}, where $S(C)\in \CP(C)$ is the image of $\mathcal{U}_C\in\CO(\Sg(C))$ under the isomorphism $\CO(\Sg(C))\raw \CP(C)$ just mentioned. Similarly, for $U\in\CO(\CA)$, the frame $\CO(\Sg_U)$ may be identified with maps $S:U \raw \CP(A)$ satisfying the conditions in \er{eerste}.
\section{External Gelfand spectrum}
It is not so easy for C*-algebraists to deal with pointfree spaces in a sheaf topos $\Sh(X)$. Fortunately, such spaces have a so-called \emph{external description} in ordinary set theory \cite{FourmanScott, johnstone02b,joyaltierney84}. In fact, a pointfree space 
$\underline{Y}$ in $\Sh(X)$ may be represented by a continuous map $\pi:Y\raw X$, where $Y$ is a pointfree space in the usual sense (i.e., in \Sets), with frame $\CO(Y)=\CO(\underline{Y})(X)$; here $\CO(\underline{Y})$ is the internal frame in  $\Sh(X)$ associated to $\underline{Y}$. The reader will now have gotten used to the idea that the notation $\pi:Y\raw X$ really denotes a frame map $\pi^*:\CO(X)\raw\CO(Y)$, nothing being implied about the possible spatiality of the frames in question. In terms of $\pi^*$, one may reconstruct $\underline{Y}$ from $\pi:Y\raw X$ as the sheaf 
\begin{equation}
\CO(\underline{Y}): U\mapsto \{V\in\CO(Y)\mid V\leq\pi^*(U)\},\:\: U\in\CO(X). \label{PTJ}
\end{equation}

Furthermore, if 
$\underline{Y}_1$ and $\underline{Y}_2$ are two pointfree spaces in $\Sh(X)$, with external descriptions
$\pi_i:Y_i\raw X$, $i=1,2$, then an internal continuous map $\underline{f}: \underline{Y}_1\raw \underline{Y}_2$ is given externally by a continuous map $f:Y_1\raw Y_2$ satisfying $\pi_2\circ f=\pi_1$.

Applying this to $X=\CA$ and $Y=\Sg$ we obtain:
\begin{theorem}\label{t2}
The  external description of the pointfree Gelfand spectrum $\underline{\Sg}$ may be identified with the canonical  projection\footnote{ That is, if $\sg\in\Sg(C)\subset\Sg$, then $\pi(\sg)=C$. From this point of view, $\CO(\Sg)$ is actually the weakest topology making this projection continuous with respect to the Alexandrov topology on $\CA$. } 
\begin{equation}
\pi:\Sg\raw\CA, \label{EGS}
\end{equation}
where $\Sg$ is seen as an ordinary (rather than a pointfree) topological space, as is $\CA$.
\end{theorem}
Taking $X=\CA$ and 
 $\underline{Y}=\underline{\Sg}$, we see from \er{sander} that
$\CO(\underline{\Sg})(\CA)=\CO(\Sg)$,
which frame is obviously spatial.\footnote{To be precise, in pointfree topology a notation like \er{EGS} is typically used for a map between pointfree spaces, which by definition is the frame map $\pi\inv: \CO(\CA)\raw\CO(\Sg)$. In this case, however, the frame map $\pi\inv$ is actually the inverse image map of the continuous map \er{EGS}, interpreted in the usual topological way.} Conversely, from \er{PTJ} and \er{EGS} we  immediately recover \er{sander}. \hfill Q.E.D.

\smallskip

Theorem \ref{t2}  has a number of interesting applications. 
We first turn to the Gelfand transform.\footnote{Unlike other approaches to Gelfand duality for noncommutative C*-algebras, our aim is not to reconstruct $A$,
but rather its `Bohrification' $\uA$, since it is the latter that carries the physical content of $A$ (at least, according to Niels Bohr's `doctrine of classical concepts' \cite{bohr49} as reformulated mathematically in \cite{landsman07}).}
The Gelfand isomorphism \er{CGD} holds internally in $\Sh(\CA)$, i.e., one has
 \begin{equation}
\uA\cong C(\underline{\Sg}, \underline{\C}) \label{CGD2}
\end{equation}
as an isomorphism of sheaves respecting the C*-algebraic structure on both sides.\footnote{Recall that isomorphisms of sheaves in sheaf toposes are simply natural isomorphisms of functors \cite{maclanemoerdijk92}. } Here $\underline{\C}$ is the pointfree space of complex numbers in $\Sh(\CA)$ with associated frame $\CO(\underline{\C})$,\footnote{Not to be confused with the \emph{complex numbers object} in $\Sh(\CA)$, given by the sheaf $U\mapsto C(U,\C)$.} defined by the  sheaf
\begin{equation}
\CO(\underline{\C}): U \mapsto \CO(U\x\C), \:\: U\in\CO(\CA).
\end{equation}
It follows from eq.\ (5.12) in \cite[Sect.\ 5]{CHLS} and \er{sander} that as a sheaf  one has
\begin{equation}
C(\underline{\Sg}, \underline{\C}):U \mapsto C(\Sg_{U},\C), \label{object}
\end{equation}
where $\Sg_U=\coprod_{C\in U} \Sg(C)$; if $U\subseteq V$, the map $C(\Sg_{V},\C)\raw C(\Sg_{U},\C)$
is given by the pullback of the inclusion $\Sg_U\hookrightarrow\Sg_V$ (that is, by restriction).
It then follows from
\er{Bohr} and \er{object} that  the isomorphism \er{CGD2} is given by its components
\begin{equation}
\uA(U)\cong C(\Sg_{U},\C).
\end{equation}
In particular, the component of the natural isomorphism in \er{CGD2} at $U=\mathop{\uparrow}\! C$ is 
\begin{equation}
C\cong C(\Sg_{\mathop{\uparrow}\! C},\C). \label{GC}
\end{equation}
A glance at the topology of $\Sg$ shows that the Hausdorffication  \er{Ha} is given by
$\Sg_{\mathop{\uparrow}\! C}^H\cong \Sg(C)$, so that the isomorphism \er{GC}
comes down to the usual Gelfand isomorphism
\begin{equation}
C\cong C(\Sg_C,\C). \label{GC2}
\end{equation}
At the end of the day, the Gelfand isomorphism \er{CGD2} therefore turns out to simply assemble all isomorphisms \er{GC2} for the commutative C*-subalgebras $C$ of $A$ into a single sheaf-theoretic construction.
 Incidentally, taking $C=\C\cdot 1$  in \er{GC} shows that $\Sg^H$ is a point, which is  also obvious from the fact that any open set containing the point $\Sg(\C\cdot 1)$ of $\Sg$ must be all of $\Sg$.
 
 \medskip
 
Second, we give a topological reinterpretation of the celebrated Kochen--Specker Theorem \cite{kochenspecker67}.\footnote{It was the sheaf-theoretic reformulation of the Kochen--Specker Theorem  by Butterfield and Isham \cite{butterfieldisham1} that originally got the
the use of topos theory in the foundations of quantum physics going. What follows is a  simplification of Sect. 6 in \cite{CHLS}, at which time the spatial nature of $\Sg$ was not yet understood.
See also \cite[Theorem 6]{HLS} for an internal proof of the equivalence between the first two bullet points.} 
 We say that a \emph{valuation} on a C*-algebra $A$ is a nonzero map $\lm:A_{\mathrm{sa}}\raw \R$  that is linear on commuting operators and dispersion-free, i.e., $\lm(a^2)=\lm(a)^2$ for all $a\in A_{\mathrm{sa}}$. 
 If $A$ is commutative, the Gelfand spectrum $\Sg(A)$ consists precisely of the valuations on $A$.\footnote{
  Physically, a valuation correspond to a so-called \emph{noncontextual hidden variable}, which
 assigns a sharp value to each observable $a$ \emph{per se}. A \emph{contextual hidden variable}
 gives a sharp value to $a$ \emph{seen in a specific measurement context in which it, in particular, may be measured}.  
 See, e.g., \cite{Redhead}. In our mathematization,  measurement contexts are identified with commutative C*-subalgebras of some ambient noncommutative C*-algebra $A$, so that a contextual  hidden variable assigns a value to a pair $(a,C)$ where $a\in C$. Hence Theorem \ref{t3} identifies \emph{noncontextual} hidden variables with \emph{continuous} cross-section of $\pi:  \Sg\raw \CA$, 
\emph{contextual} hidden variable corresponding to possibly \emph{discontinuous} cross-sections.

The mathematics neatly fits the physics here, but it should be realized that specific examples of C*-algebras $A$ may suggest coarser natural topologies on $\CA$ than the Alexandrov topology (like the Scott topology), which in turn may imply stronger continuity conditions. We thank the referee for this comment.} 
\begin{theorem}\label{t3}
There is a bijective correspondence between:
\begin{itemize}
\item 
Valuations on $A$;
\item Points  of $\underline{\Sg}(\uA)$  in $\Sh(\CA)$;
\item Continuous cross-sections $\sg:\CA\raw\Sg$  of the bundle $\pi:  \Sg\raw \CA$ of Theorem \ref{t2}.
\end{itemize}
In particular, this bundle admits no continuous cross-sections as soon as $A$ has no valuations,\footnote{The claim following this footnote sign  is the content of the original Kochen--Specker Theorem \cite{kochenspecker67}.} as in the case $A=B(H)$ with $\dim(H)>2$.
\end{theorem}
To prove this, we first give the external description of points of a pointfree space  $\underline{Y}$ in a sheaf topos $\Sh(X)$.
The subobject classifier in $\Sh(X)$ is the sheaf $\underline{\Omega}:U\mapsto \CO(U)$, in terms of which
a point of $\underline{Y}$ is  a frame map $\CO(\underline{Y})\raw \underline{\Omega}$.  
Externally, the pointfree space defined by the frame $\underline{\Omega}$ is given by the identity map $\mathrm{id}_X:X\raw X$, so that
a point of  $\underline{Y}$ externally correspond to a continuous cross-section  $\sg:X\raw Y$ of the bundle $\pi:Y\raw X$ (i.e., $\pi\circ\sg=\mathrm{id}_X$).
 In principle, $\pi$ and $\sg$  are by definition frame maps in the opposite direction, but in the case  at hand, namely $X=\CA$ and $Y=\Sg$, the map  $\sg:\CA\raw\Sg$ may be interpreted as a  continuous cross-section of the projection \er{EGS} in the usual sense. 
 Being a cross-section simply means that $\sg(C)\in\Sg(C)$. As to continuity,
by definition of the Alexandrov topology, $\sg$ is continuous iff the following condition is satisfied:
\begin{quote}
for all $\mathcal{U}\in\CO(\Sg)$
and all $C\subseteq D$, if $\sg(C)\in\mathcal{U}$ then $\sg(D)\in\mathcal{U}$.
\end{quote}  Hence, given the definition of $\CO(\Sg)$, the following condition is sufficient for continuity: if  $C\subseteq D$, then $\sg(D)_{|C}=\sg(C)$. However, this condition is also necessary. To explain this, let $\rh_{DC}:\Sg(D)\raw\Sg(C)$ again be the restriction map. This map is continuous and open. 
Suppose $\rh_{DC}(\sg(D))\neq \sg(C)$.  Since $\Sg(D)$ is Hausdorff, there is an open neighbourhood $\mathcal{U}_D$ of $\rh_{DC}\inv(\sg(C))$ not containing $\sg(D)$. Let $\mathcal{U}_C=\rh_{DC}(\mathcal{U}_D)$ and take any $\mathcal{U}\in\CO(\Sg)$ such that
$\mathcal{U}\cap\CO(\Sg(C))=\mathcal{U}_C$ and $\mathcal{U}\cap\CO(\Sg(D))=\mathcal{U}_D$. This is possible, since 
$\mathcal{U}_C$ and $\mathcal{U}_D$ satisfy both conditions in the definition of $\CO(\Sg)$. By construction,
$\sg(C)\in \mathcal{U}$ but $\sg(D)\notin \mathcal{U}$, so that $\sg$ is not continuous. 
Hence $\sg$ is a continuous cross-section of $\pi$ iff 
\begin{equation}
\sg(D)_{|C}=\sg(C) \mbox{ for all } C\subseteq D. \label{sgcon}
\end{equation}
Now define a map $\lm:A_{\mathrm{sa}}\raw\C$ by $\lm(a)=\sg(C^*(a))(a)$, where 
$C^*(a)$ is the commutative  unital C*-algebra generated by $a$. If $b^*=b$ and $[a,b]=0$, then $\lm(a+b)=\lm(a)+\lm(b)$ 
by \er{sgcon}, applied to $C^*(a)\subset C^*(a,b)$ as well as to $C^*(b)\subset C^*(a,b)$. Furthermore, since $\sg(C)\in\Sg(C)$,
the map $\lm$ is dispersion-free. 

Conversely, a valuation $\lm$ defines a cross-section $\sg$ by complex linear extension of $\sg(C)(a)=\lm(a)$, where $a\in C_{\mathrm{sa}}$. By the criterion \er{sgcon} this cross-section is evidently continuous, since the value $\lm(a)$ is independent of the choice of $C$ containing $a$.
 \hfill Q.E.D.

\medskip

The contrast between the pointlessness of the internal spectrum $\ulS$ and the spatiality of the external spectrum $\Sg$ is quite striking, but easily explained:  a point of  $\Sg$ (in the usual sense, but also in the frame-theoretic sense  in the case that $\Sg$ is sober)
necessarily lies in some $\Sg(C)\subset\Sg$, and hence is defined (and dispersion-free) only in the `context' $C$.
 For example, for $A=M_n(\C)$, a point $\lm\in\Sg(C)$ corresponds to a map
 \begin{equation}
  \lm^*:\CO(\Sg)\raw\{0,1\}, \:\:
S\mapsto \lm(S(C)),
\end{equation}
where $\CO(\Sg)$ has been realized as in \er{eerste}. Thus $\lm^*$ is only sensitive to the value of $S$ at $C$.

\medskip

To close, we examine the possible soberness of $\Sg$ \cite[Theorem 8]{S2010}, \cite[Theorem 2.25]{SW}:
\begin{proposition}\label{t4}
The space $\Sg$ is sober if $A$ satisfies the \emph{ascending chain condition}: every chain $C_1\subseteq C_2\subseteq \cdots$ of elements $C_i\in\CA$ converges, in  that $C_{n}=C_m$ for all $n>m$.
\end{proposition}
The proof is straightforward, relying on the identification of points of $\Sg$ with irreducible closed subsets of $S$ and the ensuing condition that $\Sg$ is sober iff every irreducible closed subsets of $S$ is the closure of a unique point \cite[\S IX.3]{maclanemoerdijk92}.

For example, this proposition implies that $\Sg$ is sober for $A=M_n(\C)$, and, more generally, for all finite-dimensional C*-algebras. 

{\small
\bibliographystyle{plain}
\bibliography{carey}

\begin{thebibliography}{10}

\bibitem{Akemann}
Charles~A. Akemann.
\newblock A {G}elfand representation theory for {$C^{\ast} $}-algebras.
\newblock {\em Pacific Journal of Mathematics}, 39:1--11, 1971.

\bibitem{banaschewskimulvey00b}
Bernhard Banaschewski and Christopher~J. Mulvey.
\newblock The spectral theory of commutative {C}*-algebras: the constructive
  {G}elfand-{M}azur theorem.
\newblock {\em Quaestiones Mathematicae}, 23:465--488, 2000.

\bibitem{banaschewskimulvey00a}
Bernhard Banaschewski and Christopher~J. Mulvey.
\newblock The spectral theory of commutative {C}*-algebras: the constructive
  spectrum.
\newblock {\em Quaestiones Mathematicae}, 23:425--464, 2000.

\bibitem{banaschewskimulvey06}
Bernhard Banaschewski and Christopher~J. Mulvey.
\newblock A globalisation of the {G}elfand duality theorem.
\newblock {\em Annals of Pure and Applied Logic}, 137:62--103, 2006.

\bibitem{BBH}
Benno~{van den} {B}erg and Chris Heunen.
\newblock Noncommutativity as a colimit.
\newblock {\em Under consideration for Applied Categorical Structures}, 2010.
\newblock arXiv:1003.3618v2.

\bibitem{bohr49}
Niels Bohr.
\newblock Discussion with {E}instein on epistemological problems in atomic
  physics.
\newblock In {\em Albert Einstein: Philosopher-Scientist}, pages 201--241. La
  Salle: Open Court, 1949.

\bibitem{borceux3}
Francis Borceux.
\newblock {\em Handbook of categorical algebra. 3. Categories of sheaves}.
\newblock Cambridge University Press, 1994.

\bibitem{butterfieldisham1}
Jeremy Butterfield and Chris~J. Isham.
\newblock A topos perspective on the {K}ochen-{S}pecker theorem: {I}. quantum
  states as generalized valuations.
\newblock {\em International Journal of Theoretical Physics}, 37:2669--2733,
  1998.

\bibitem{CHLS}
Martijn Caspers, Chris Heunen, Nicolaas~P. Landsman, and Bas Spitters.
\newblock Intuitionistic quantum logic of an {$n$}-level system.
\newblock {\em Foundations of Physics}, 39:731--759, 2009.

\bibitem{CIR}
Rachid Choukri, El~Hossein Illoussamen, and Volker Runde.
\newblock Gelfand theory for non-commutative {B}anach algebras.
\newblock {\em The Quarterly Journal of Mathematics}, 53:161--172, 2002.

\bibitem{Connes}
Alain Connes.
\newblock {\em Noncommutative geometry}.
\newblock Academic Press Inc., San Diego, CA, 1994.

\bibitem{ConnesMarcolli}
Alain Connes and Matilde Marcolli.
\newblock {\em Noncommutative geometry, quantum fields and motives}.
\newblock American Mathematical Society, Providence, RI, 2008.

\bibitem{coquand05}
Thierry Coquand.
\newblock About {S}tone's notion of spectrum.
\newblock {\em Journal of Pure and Applied Algebra}, 197:141--158, 2005.

\bibitem{CS}
Thierry Coquand and Bas Spitters.
\newblock Constructive {G}elfand duality for {$C^*$}-algebras.
\newblock {\em Mathematical Proceedings of the Cambridge Philosophical
  Society}, 147:339--344, 2009.

\bibitem{DaunsHofmann}
John Dauns and Karl~Heinrich Hofmann.
\newblock {\em Representation of rings by sections}.
\newblock Memoirs of the American Mathematical Society, No. 83. American
  Mathematical Society, Providence, R.I., 1968.

\bibitem{FellDoran2}
J.~M.~G. Fell and R.~S. Doran.
\newblock {\em Representations of {$^*$}-algebras, locally compact groups, and
  {B}anach {$^*$}-algebraic bundles. {V}ol. 2}.
\newblock Academic Press Inc., Boston, MA, 1988.

\bibitem{FourmanScott}
M.~P. Fourman and D.~S. Scott.
\newblock Sheaves and logic.
\newblock In {\em Applications of sheaves ({P}roc. {R}es. {S}ympos. {A}ppl.
  {S}heaf {T}heory to {L}ogic, {A}lgebra and {A}nal., {U}niv. {D}urham,
  {D}urham, 1977)}, volume 753 of {\em Lecture Notes in Math.}, pages 302--401.
  Springer, Berlin, 1979.

\bibitem{goldblatt84}
Robert Goldblatt.
\newblock {\em Topoi, the categorical analysis of logic}.
\newblock North-Holland, 1984.

\bibitem{HLS}
Chris Heunen, Nicolaas~P. Landsman, and Bas Spitters.
\newblock A topos for algebraic quantum theory'.
\newblock {\em Communications in Mathematical Physics}, 291:63--110, 2009.

\bibitem{johnstone82}
Peter~T. Johnstone.
\newblock {\em Stone Spaces}.
\newblock Cambridge University Press, 1982.

\bibitem{johnstone02a}
Peter~T. Johnstone.
\newblock {\em Sketches of an Elephant: A topos theory compendium}, volume~1.
\newblock Clarendon Press, 2002.

\bibitem{johnstone02b}
Peter~T. Johnstone.
\newblock {\em Sketches of an Elephant: A topos theory compendium}, volume~2.
\newblock Clarendon Press, 2002.

\bibitem{joyaltierney84}
Andr{\'e} Joyal and Miles Tierney.
\newblock {\em An extension of the Galois theory of Grothendieck}, volume~51.
\newblock Memoirs of the American Mathematical Society, 1984.

\bibitem{kochenspecker67}
Simon Kochen and Ernst Specker.
\newblock The problem of hidden variables in quantum mechanics.
\newblock {\em Journal of Mathematics and Mechanics}, 17:59--87, 1967.

\bibitem{Woro}
Pawe{\l} Kruszy{\'n}ski and Stanis{\l}aw~L. Woronowicz.
\newblock A noncommutative {G}elfand-{N}aimark theorem.
\newblock {\em Journal of Operator Theory}, 8:361--389, 1982.

\bibitem{PSTP}
Nicolaas~P. Landsman.
\newblock Poisson spaces with a transition probability.
\newblock {\em Reviews in Mathematical Physics}, 9(1):29--57, 1997.

\bibitem{landsman98}
Nicolaas~P. Landsman.
\newblock {\em Mathematical topics between classical and quantum mechanics}.
\newblock Springer, 1998.

\bibitem{landsman07}
Nicolaas~P. Landsman.
\newblock Between classical and quantum.
\newblock In J.~Earman J.~Butterfield, editor, {\em Handbook of Philosophy of
  Science}, volume 2: Philosophy of Physics, pages 417--553. Elsevier, 2007.

\bibitem{maclanemoerdijk92}
Saunders {Mac Lane} and Ieke Moerdijk.
\newblock {\em Sheaves in Geometry and Logic}.
\newblock Springer, 1992.

\bibitem{Mulvey80}
Christopher~J. Mulvey.
\newblock A noncommutative {G}elfand-{N}aimark theorem.
\newblock {\em Mathematical Proceedings of the Cambridge Philosophical
  Society}, 88:425--428, 1980.

\bibitem{Redhead}
Michael~L.G. Redhead.
\newblock {\em Incompleteness, nonlocality and realism: A prolegomenon to the
  philosophy of quantum mechanics}.
\newblock Clarendon Press, Oxford, 1987.

\bibitem{Shultz}
Frederic~W. Shultz.
\newblock Pure states as a dual object for {$C^{\ast} $}-algebras.
\newblock {\em Communications in Mathematical Physics}, 82:497--509, 1981/82.

\bibitem{S2010}
Bas Spitters.
\newblock The space of measurement outcomes as a spectrum for non-commutative
  algebras.
\newblock {\em EPTCS}, 2010.
\newblock doi: \texttt http://dx.doi.org/10.4204/EPTCS.26.12.

\bibitem{Vic:LocTopSp}
Steven Vickers.
\newblock Locales and toposes as spaces.
\newblock In Marco Aiello, Ian~E. Pratt-Hartmann, and Johan~F.A.K. van Benthem,
  editors, {\em Handbook of Spatial Logics}, chapter~8. Springer, 2007.

\bibitem{SW}
Sander Wolters.
\newblock Contravariant vs covariant quantum logic: A comparison of two
  topos-theoretic approaches to quantum theory.
\newblock {\em To appear}, 2010.

\end{thebibliography}
}
\end{document}